\documentclass[aps,prl,showpacs,twocolumn,superscriptaddress]{revtex4}
\usepackage{amsmath,graphics,color}
\usepackage[next]{inputenc}
\usepackage[dvips]{epsfig}
\def\be{\begin{equation}}
\def\ee{\end{equation}}

\def\bi{\begin{itemize}}
\def\ei{\end{itemize}}
\def\bn{\begin{enumerate}}
\def\en{\end{enumerate}}
\def\bea{\begin{eqnarray}}
\def\eea{\end{eqnarray}}
\def\no{\nonumber}
\def\ba{\begin{array}}
\def\ea{\end{array}}
\def\bd{\begin{displaymath}}
\def\ed{\end{displaymath}}

\begin{document}
\title{Three-Qubit Ground State and Thermal Entanglement of $XXZ$ Model With Dzyaloshinskii-Moriya Interaction}

\author{R. Jafari}
\affiliation{Institute for Advanced Studies in Basic Sciences,
Zanjan 45195-1159, Iran}

\author{A. Langari}
\affiliation{Physics Department, Sharif University of Technology,
Tehran 11155-9161, Iran}

\begin{abstract}
We have studied the symmetric and non-symmetric
pairwise ground state and thermal entanglement in three-qubit
anisotropic Heisenberg (XXZ) and Ising in a magnetic field models and in the presence of
Dzyaloshinskii-Moriya (DM) interaction. We have found that
increasing of the DM interaction and magnetic field can enhance
and reduce the entanglement of system. We have shown that the non-symmetric
pairwise has higher value concurrence and critical temperature
(above which the entanglement vanishes) than the symmetric
pairwise. For negative anisotropy the non-symmetric entanglement is
a monotonic function of DM interaction while for positive anisotropy
it has a maximum versus DM parameter and vanishes for larger values of DM interaction.
The conditions for the existence of thermal entanglement
are discussed in details. The most remarkable result happens at
zero temperature where 3-qubit ground state entanglement of the
system (in spite of 2-qubit counterpart) shows the fingerprint of
quantum phase transition for an infinite size system.
\end{abstract}
\date{\today}


\maketitle
\section{Introduction}
Entanglement is a  property of quantum state which has been studied
intensively in recent years  as a specific nonlocal quantum mechanical
correlation\cite{Schrodinger,Einstein,Bell} and it
becomes recently a key feature of quantum information
theory\cite{Bennet1}. When the entanglement is generated,
observation or manipulation of it in practice constitutes a major
obstacle, because of the fragility of quantum entanglement to the
decoherence induced by environment.
Therefore, how to generate,
maintain and control the entanglement in the presence of
dissipative coupling of the system to the environment is of utmost
importance in the implementation of quantum information
processing.

However, similar to the superposition of two coherent electromagnetic
waves which enable us to learn some global information from a
localized spatial area, people expect that the entanglement, which
roots in the same superposition principle, can enable us to learn
some global properties from a small part of the system. This
observation may be one of the main motivations in the recent
studies \cite{Gu,Larsson,Legeza} on the role of entanglement
between a small part, e.g. a block consisting of one or more
sites, and the rest of the system in the quantum phase transition.
These results suggested that the local entanglement may be used as
a good marker of quantum phase transition.

Then over the past few years there has been an ongoing effort to
characterize the entanglement properties of condensed matter
systems and apply them in quantum information. The quantum
entanglement in solid state systems such as spin chains is an
important emerging
field\cite{Nielsen,Wang1,Kamta,OConnor,Sun,Khveshchenko,Zhang1,Jafari,kargarian}.
Spin chain are natural candidates for realization of entanglement,
and spin effects have been investigated in many other systems,
such as superconductors\cite{Nishiyama}, quantum
dots\cite{Trauzettel} and trapped ions\cite{Porras}. A most known
models in the spin chains is Heisenberg model and Ising model as a
special case of Heisenberg model. The Heisenberg model can
describe interaction of qubits not only in solid physical systems
but also in many other systems such as quantum dots\cite{Loss},
nuclear spin\cite{Kane}, cavity QED\cite{Imamog-lu,Zheng}, optical
lattice\cite{Anders}, quantum computation\cite{Lidar} and
controlled-Not gate\cite{Zheng}.

In recent years the two-qubit thermal entanglement which includes
spin-spin interactions\cite{Kamta,Sun,Zhang1,Abliz,Dominic,Souza},
and spin-orbit
coupling\cite{Kheirandish,Li,Zhang2,Chuang,Gurkan,Wang2}
(Dzyaloshinskii\cite{Dzyaloshinskii}-Moriya\cite{Moriya}
interaction) has been studied. Entanglement in two-qubit state
has been well studied in the literature along various kind of
three-qubit entanglement
states\cite{Dur,Coffman,Brun,Acin,Rajagopal}. The three-qubit
entanglement states have been shown to possess advantage over the
two-qubit states in quantum teleportation\cite{Karlsson}, dense
coding\cite{Hao} and quantum cloning\cite{Brub}.
More specificly, both quantitative and qualitative behavior of the entanglement
in two-qubit and three-qubit systems are different. The concurrence is always increasing
in a two-qubit model versus DM parameter while it is both increasing and decreasing versus DM
for a three-qubit model.The position where the concurrence is maximum or zero for a three-qubit
model corresponds to the quantum critical point of the infinite size system while it is not
the case for a two-qubit model. Moreover, the pairwise entanglement can be defined
as symmetric and non-symmetric ones with differente properties for three-qubit as will be
discussed in this article.

In addition to the above facts, recently some novel magnetic systems with antiferromagnetic (AF) properties, such as $Cu(C_{6}D_{5}COO)_{2}3D_{2}O$\cite{Dender1},
$Yb_{4}As_{3}$\cite{Kohgi}, $BaCu_{2}Si_{2}O_{7}$\cite{Tsukada},
$\alpha-Fe_{2}O_{3}$, $LaMnO_{3}$\cite{Grande} and
$K_{2}V_{3}O_{8}$\cite{Greven}, were discovered in the category of
quasi-one dimensional materials which are known to belong to an
antisymmetric interaction of the form
$\overrightarrow{D}.(\overrightarrow{S_{i}}\times\overrightarrow{S_{j}})$
which is known as the Dzyaloshinskii-Moriya
(DM)interaction\cite{Dzyaloshinskii,Moriya}. Thus, investigation of the
quantum effects in such spin models requires more research in this direction.

In this paper, we have investigated the influence of the the
anisotropy coupling, magnetic field and the $z$-component DM
interaction on the non-symmetric ($\rho_{12}=Tr_{3}\rho(T)$) and
symmetric pairwise ($\rho_{13}=Tr_{2}\rho(T)$) ground state and
thermal entanglement of the three-qubit $XXZ$ and Ising models in a
magnetic field and in the presence of DM interaction. We will show
that the DM interaction, anisotropy and magnetic field parameters
are efficient control parameters of entanglement. Increasing the
DM coupling and anisotropy have a different effects on the
entanglement and can enhance or reduce the entanglement, whereas
these parameter just increase the entanglement in two-qubit
\cite{Kheirandish,Li,Zhang2,Chuang,Gurkan,Wang2} counterpart. However
we show that at $T=0$ the 3-qubit quantum phase transition points
of this system correspond to those ones at thermodynamic
limit\cite{Alcaraz,Jafari}.

\section{The model interaction}

The Hamiltonian of N-qubit $XXZ$ model with $z$-component
of DM Interaction is

\bea \label{eq4}
H(\tilde{J},\Delta)=\frac{\tilde{J}}{4}\sum_{i}^{N}\Big[\sigma_{i}^{x}\sigma_{i+1}^{x}+\sigma_{i}^{y}\sigma_{i+1}^{y}+
\Delta\sigma_{i}^{z}\sigma_{i+1}^{z}\\
\no
+D(\sigma_{i}^{x}\sigma_{i+1}^{y}-\sigma_{i}^{y}\sigma_{i+1}^{x})\Big],
\eea
where  $\tilde{J}$ is the exchange coupling, $D$ is the strength of $z$ component
DM interaction and $\Delta$ defines the easy-axis anisotropy
which can be positive or negative. The positive or negative $\tilde{J}$ corresponds to the antiferromagnetic (AF) or
ferromagnetic (F) cases, respectively. $\sigma_{i}^{\alpha}$ refers to the $\alpha$-component
of Pauli matrix at site $i$. A $\pi$-rotation around $z$ axis on odd (or even) sites maps the F case ($\tilde{J}<0$)
to the AF case with the opposite sign of anisotropy,
\bea \label{eq5}
H(J,\Delta)=\frac{J}{4}\sum_{i}^{N}\Big[\sigma_{i}^{x}\sigma_{i+1}^{x}+\sigma_{i}^{y}\sigma_{i+1}^{y}-
\Delta\sigma_{i}^{z}\sigma_{i+1}^{z}\\
\no
+D(\sigma_{i}^{x}\sigma_{i+1}^{y}-\sigma_{i}^{y}\sigma_{i+1}^{x})\Big],~~J=|\tilde{J}|>0.
\eea
So we can restrict ourselves to AF case ($J>0$) with $D>0$ and arbitrary anisotropy ($\Delta<0$ and
$\Delta>0$) without loss of generality.

We can also restore the Ising model with DM interaction
(IDM) in the limit $J\rightarrow 0, \Delta
\rightarrow \infty \; \mbox{and} \; D \rightarrow \infty$ where $
J\Delta=\hat{J} \; \mbox{and} \; \frac{D}{\Delta}=\hat{D}$. The
resulting Hamiltonian is given by \bea \label{eq5-b}
H(\hat{J},\hat{D})=\frac{\hat{J}}{4}\sum_{i}^{N}
\Big[\sigma_{i}^{z}\sigma_{i+1}^{z}
+\hat{D}(\sigma_{i}^{x}\sigma_{i+1}^{y}-\sigma_{i}^{y}\sigma_{i+1}^{x})\Big].
\eea It should be noticed that both AF (corresponds
to the positive anisotropy in $XXZ$ with $DM$) and F cases
(corresponds to the negative anisotropy in $XXZ$
with $DM$) have to be considered separately, since the mentioned
symmetry can not be justified for Eq.(\ref{eq5-b}). We will
consider the above Hamiltonians (Eqs.(\ref{eq5}, \ref{eq5-b})) on
a string of 3-qubits and obtain the entanglement of them in terms
of the model parameters.

The entanglement of two-qubit can be measured by the concurrence which is defined as\cite{Wootters}
\bea \label{eq1}
C_{ij}=\max\Big\{2\max(\lambda_{k})-\sum_{k=1}^{4}\lambda_{k},0\Big\}
\eea
where $\lambda_{k}$ (k=1,2,3,4) are the square roots of the
eigenvalues of the following operator
\bea \label{eq2}
R=\rho_{ij}(\sigma_{i}^{y}\otimes\sigma_{j}^{y})\rho_{ij}^{\ast}(\sigma_{i}^{y}\otimes\sigma_{j}^{y}),
\eea
$\rho_{ij}$ is the density matrix of pair $i$ and $j$ spins and asterisk denotes the complex conjugate.
For a system in equilibrium at temperature $T$ the state of a system is determined by the
density matrix
\bea \label{eq3} \rho(T)=\frac{e^{-\beta H}}{Z}, \eea
where $H$ is
the system Hamiltonian, $Z=Tr(e^{-\beta H})$ is the partition function and $\beta=\frac{1}{K_{B}T}$ where $T$ is
temperature and $K_{B}$ is Boltzmann constant. For simplicity we take $K_{B}=1$.

To get the concurrence of two qubits in a string of 3-qubits we define two types of
reduced density matrix, the symmetric and non-symmetric ones. Let label the 3-qubits
as $1, 2, 3$ sequentially. The symmetric reduced density matrix ($\rho_{13}$)
is defined as $\rho_{13}=tr_{2} (\rho)$, where $\rho$ is the density matrix of
3-qubits. The non-symmetric reduced density matrix  is  $\rho_{12}=tr_{3} (\rho)$.

\begin{figure}
\begin{center}
\includegraphics[width=6cm]{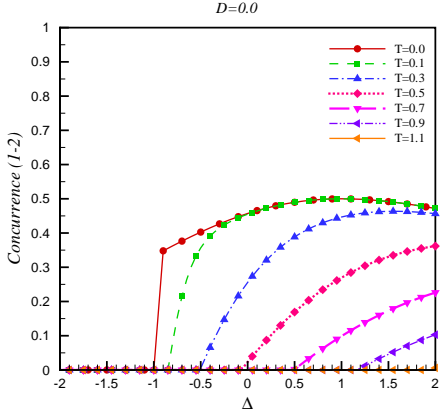}
\caption{(color online) The concurrence of non-symmetric pairwise is
plotted versus $\Delta$ for different values of $T$, ($D=0.0$).} \label{fig1}
\end{center}
\end{figure}

\begin{figure}[<t>]
\begin{center}
\includegraphics[width=6cm]{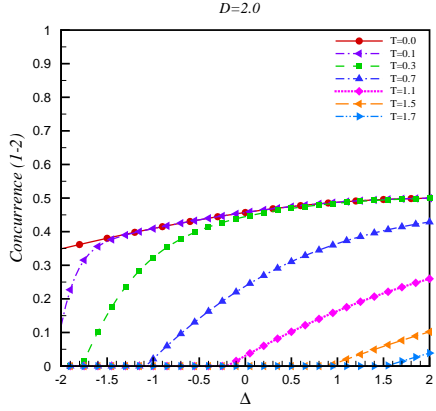}
\caption{(color online) The concurrence of non-symmetric pairwise
against anisotropy ($\Delta$) for different values of $T$
,($D=2.0$).} \label{fig2}
\end{center}
\end{figure}

\section{Three qubits XXZ with DM interaction}

A straightforward calculation gives the eigenstates and eigenvalues of the 3-qubit XXZ+DM Hamiltonian \cite{Jafari}.
The positive square roots of the eigenvalues of the matrix
$R$ for non-symmetric pairwise is obtained
\bea \no
\lambda_{1}&=&\lambda_{2}=\frac{1}{Z}\Big\{\frac{1}{2}+e^{-\frac{J}{2}\beta\Delta}\\
\no &+&e^{\frac{J}{4}\beta\Delta}
\Big[\frac{1}{2}\cosh(\frac{J}{4}\beta
q)-\frac{\Delta}{2q}\sinh(\frac{J}{4}\beta q)\Big]\Big\},\\
\no
\lambda_{3}&=&\frac{1}{Z}\Big\{\frac{1}{2}+e^{\frac{J}{4}\beta\Delta}\times\\
\no &&\Big[\frac{3}{2}\cosh(\frac{J}{4}\beta
q)+\frac{\Delta-8\sqrt{1+D^{2}}}{2q}\sinh(\frac{J}{4}\beta q)\Big]\Big\},\\
\no
\lambda_{4}&=&\frac{1}{Z}\Big\{\frac{1}{2}+e^{\frac{J}{4}\beta\Delta}\times\\
\no&& \Big[\frac{3}{2}\cosh(\frac{J}{4}\beta
q)+\frac{\Delta+8\sqrt{1+D^{2}}}{2q}\sinh(\frac{J}{4}\beta
q)\Big]\Big\}.\\ \label{eqC1}\eea

The partition function is then
\bea \label{eqC2}
Z&=&4e^{\frac{J}{4}\beta\Delta}\cosh(\frac{J}{4}\beta q)+
4e^{-\frac{J}{4}\beta\Delta}\cosh(\frac{J}{4}\beta\Delta)
 \eea
The square root of matrix $R$ is invariant under the substitution $D
\rightarrow -D$, so the generality of AF case with $D>0$ and arbitrary anisotropy, remains pristine.

\begin{figure}
\begin{center}
\includegraphics[width=6cm]{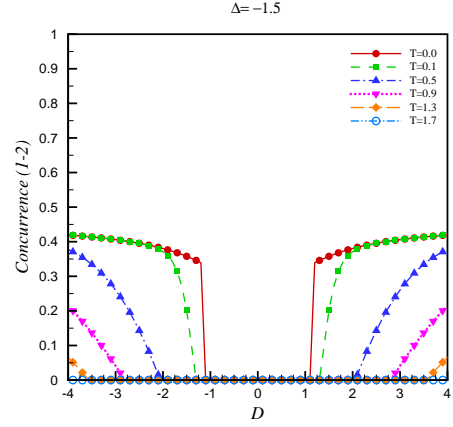}
\caption{(color online) The variation of non-symmetric pairwise
concurrence with DM coupling ($D$) for different values of $T$,
$\Delta=-1.5$.} \label{fig3}
\end{center}
\end{figure}

    At $T=0$, for $\Delta > -\sqrt{1+D^{2}}$ the concurrence of the doubly-degenerate ground state is
    \bea
    \no
    C_{12}(T=0)=\frac{2(q+\Delta-\sqrt{1+D^{2}})\sqrt{1+D^{2}}}{q(q+\Delta)},
    \eea

    while it is a disentangled ferromagnetic states for $\Delta<-\sqrt{1+8D^{2}}$. At the level crossing line ($\Delta=-\sqrt{1+8D^{2}}$) the ground state is 4-fold degenerate and disentangled.

    For nonzero temperature ($T\neq0$) the concurrence is
    \begin{eqnarray}
    \no
    C_{12}^{AF}(T)=\frac{e^{\frac{J}{4}\beta}}{Z}\max\Big\{(\frac{1+8D}{q})\sinh(\frac{J}{4}\beta
    q)\\
    \no
    -\cosh(\frac{J}{4}\beta q)
    -e^{-\frac{J}{4}\beta}-2e^{-\frac{3J}{4}\beta},0\Big\}.
    \end{eqnarray}

    The model is entangled if $T>T_{c}(\Delta,D)$, and $T_{c}(\Delta,D)$ is determined by the
    following nonlinear equation
    \bea
    \label{eq7}
    (\frac{1+8D}{q})\sinh(\frac{J}{4}\beta q)
    -\cosh(\frac{J}{4}\beta q)=e^{-\frac{J}{4}\beta}+2e^{-\frac{3J}{4}\beta}.
     \eea

\begin{figure}
\begin{center}
\includegraphics[width=6cm]{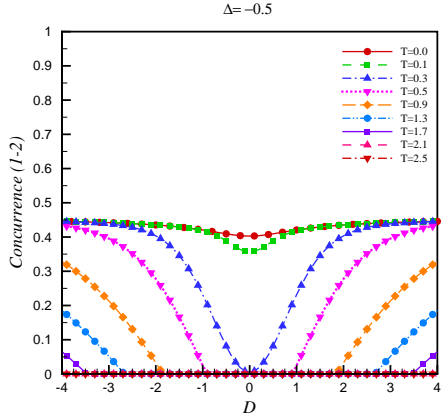}
\caption{(color online) The variations of non-symmetric pairwise
concurrence as a function of DM coupling ($D$) for different values of $T$,
$\Delta=-0.5$.} \label{fig4}
\end{center}
\end{figure}

    To determine whether the entanglement exists or not, we have to consider two different cases.
    \begin{itemize}
    \item $\Delta<0$\\
    The ground state entanglement of non-symmetric pairwise have been shown in
    Fig.(\ref{fig1}) and Fig.(\ref{fig2})  as a function of $\Delta$ for $D=0$ and $D=2$
which are denoted by $T=0$ plots.
Both figures manifest that the concurrence of ground state decreases by reduction of the anisotropy and
suddenly become zero below the critical line $\Delta_{c}=-\sqrt{1+D^{2}}$.
    In the other words, for each value of $D$, there is a threshold  $\Delta_{c}=-\sqrt{1+D^{2}}$
    above which the ground state will be entangled.
    Accordingly, for each value of anisotropy ($\Delta\leq-1$) there is a value of DM
    coupling $D_{c}=\sqrt{\Delta^{2}-1}$ under which the
    ground state looses its entanglement. The critical line $\Delta_{c}=-\sqrt{1+D^{2}}$ is at the position where
the level crossing occurs and corresponds to the critical line of this model at thermodynamic
    limit (i.e. infinite number of qubits). At this critical line the global $U(1)\times Z_{2}$
    symmetry of the  Hamiltonian changes to the local $SU(2)$ symmetry.

    At finite temperature, thermal entanglement
    behaves similar to the ground state counterpart except that it becomes disentangled gradually by
increasing temperature. For $T\neq0$, the critical line below which the thermal entanglement vanishes is a function
 of $D$  and temperature, i.e $\Delta_{c}(D, T)$ and is given by Eq.(\ref{eq7}). Generally, the decrease in
anisotropy parameter and increase in temperature have a reduction influence on the concurrence. The effect
of temperature is plotted in Fig.(\ref{fig1}) and Fig.(\ref{fig2}) for some fixed values of temperature.
    To scan the influence of DM coupling the concurrence has been plotted in Fig.(\ref{fig3}) and Fig.(\ref{fig4})
at fixed values of $\Delta=-1.5$ and $\Delta=-0.5$ versus $D$.
For $\Delta<-1$, the concurrence jumps suddenly to non-zero value as $D$ crosses the critical value
 $D_{c}=\sqrt{\Delta^{2}-1}$ as shown in Fig.(\ref{fig3}). Contrary to the previous case, for $-1\leq\Delta<0$
the ground state is always entangled while the thermal entanglement becomes zero for $|D|<D_{c}(\Delta, T)$
which is justified in Fig.(\ref{fig4}).

    At $\Delta=0$ the concurrence of ground state is independent of DM coupling and take a constant value,
    \be
    \no
    C_{12}(T=0,\Delta=0)=\frac{2\sqrt{2}-1}{4}.
    \ee

\begin{figure}
\begin{center}
\includegraphics[width=6cm]{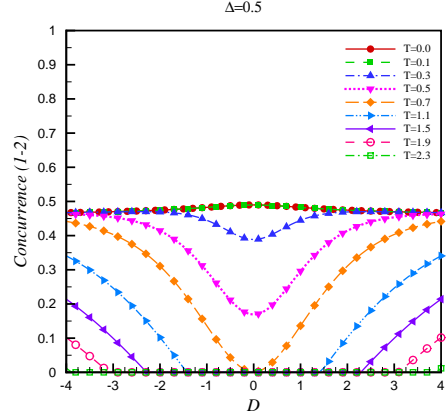}
\caption{(color online) The concurrence of non-symmetric pairwise
is plotted versus ($D$) for different values of temperature,
$\Delta=0.5$.} \label{fig5}
\end{center}
\end{figure}

\begin{figure}
\begin{center}
\includegraphics[width=6cm]{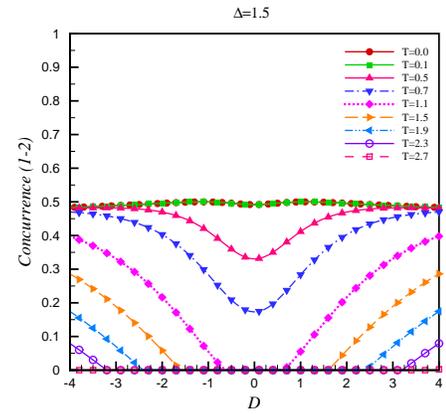}
\caption{(color online) The variations of non-symmetric pairwise
concurrence with DM coupling ($D$) for different values of $T$,
$\Delta=1.5$.} \label{fig6}
\end{center}
\end{figure}

    \item $\Delta>0$\\
    In the positive anisotropic region, increasing the anisotropy enhances the ground state and thermal entanglement
    up to a maximum and then decreases gradually which can be seen in Fig.(\ref{fig5}) and Fig.(\ref{fig6}).
At $T=0$ the maximum value of concurrence is due to the maximum fluctuations which exist at the critical line $\Delta_{c}=\sqrt{1+8D^{2}}$. The critical line belongs to the spin-fluid to N\'{e}el phase transition
for infinite number of qubits \cite{Jafari}. Contrary to what happened in the negative
    anisotropy region where the ground state entanglement vanishes immediately below the critical line
    $\Delta_{c}=-\sqrt{1+D^{2}}$, in the positive anisotropy region the ground state entanglement has non-zero
value in the both side of critical line $\Delta_{c}=\sqrt{1+D^{2}}$. However, in the both cases the global
    $U(1)\times Z_{2}$ symmetry of the Hamiltonian breaks to the local $SU(2)$ symmetry on the critical line
 ($\Delta_{c}=\mp\sqrt{1+D^{2}}$). The influence of temperature is to decrease the concurrence and
increase the critical anisotropy parameter ($\Delta_{c}(D,T)$) beyond which the system being entangled
(as given by Eq.(\ref{eq7})).
    To survey the effect of DM interaction the ground state and thermal concurrence have been plotted in
    Fig.(\ref{fig5}) and Fig.(\ref{fig6}) versus $D$ for $\Delta=0.5$, $\Delta=1.5$ and different values of
    temperature.
    As far as $0<\Delta<1$ the influence of DM
    interaction is to reduce the concurrence of ground state (Fig.(\ref{fig5})) while for $\Delta\geq1$
the ground state concurrence gets a maximum at the position of the critical line $\Delta_{c}=\sqrt{1+D^{2}}$.
In the other words, for $\Delta\geq1$ increasing of the DM interaction raises the ground state concurrence until its maximum value and then decreases slowly (Fig.(\ref{fig6})). This is a feature which is only observable
in 3-qubits system and is related to the fact that the 3-qubits system show the critical behavior of
infinite qubits limit correctly.
    For low temperatures, thermal entanglement behaves similar to the ground state concurrence, but at
high temperatures the thermal entanglement is increased on the onset of the DM coupling.

\item Symmetric pairwise\\
We have discussed extensively the non-symmetric pairwise entanglement ($C_{12}$) which has been calculated
from $\rho_{12}$. Now, we summerize the main features of the symmetric pairwise entanglement ($C_{13}$)
in comparison with the non-symmetric one without presenting the details.

In the negative anisotropy region ($\Delta<0$),
there is a threshold anisotropy and $D$ parameters beyond that the symmetric
and non-symmetric pairwise are entangled. Decreasing the anisotropy increases the
ground state entanglement of symmetric pairwise but decreases the ground state
entanglement of non-symmetric pairwise. However, reduction of the anisotropy increases the
thermal entanglement of symmetric pairwise up to a  maximum value and then decreases it
gradually to be vanished while it decreases the non-symmetric pairwise thermal concurrence
to zero. In symmetric pairwise the increment of DM coupling decreases the ground state
entanglement but enhances the thermal entanglement up to the maximum value and then decreases it.
For non-symmetric pairwise the influence of DM interaction on ground state and thermal concurrence
is incremental.
For the positive anisotropy region ($\Delta>0$), the symmetric case is disentangled in the presence or absence of
temperature.

    \end{itemize}

\section{Three qubits Ising with DM interaction}

We have studied both symmetric and non-symmetric pairwise
entanglement of the Ising model with DM interaction defined by
Hamiltonian in Eq.(\ref{eq5-b}). Its qualitative behavior is
similar to the XXZ model with DM interaction while only some
quantitative changes observed, for example the
critical point for both AF and F cases is $D_{c}=1$. Hence, we
do not present the results of 3-qubits Ising model with DM
interaction here; however, we will show the effect of magnetic
field on the entanglement in this model. Let rewrite the
Hamiltonian of 3-qubit Ising model with DM interaction in the
following form \bea \no H=\frac{J}{4}\Big[
(\sigma_{1}^{z}\sigma_{2}^{z}+\sigma_{2}^{z}\sigma_{3}^{z})+
D(\sigma_{1}^{x}\sigma_{2}^{y}-\sigma_{1}^{y}\sigma_{2}^{x})\\
+D(\sigma_{2}^{x}\sigma_{3}^{y}-\sigma_{2}^{y}\sigma_{3}^{x})+h
(\sigma_{1}^{z}+\sigma_{2}^{z}+\sigma_{3}^{z})\Big],
\label{idmh}
\eea
where $h$ is proportional to the strength of magnetic field, $J$ and $D$ are the
exchange and DM couplings, respectively.
 The square root of the eigenvalues of matrix $R$ (Eq.(\ref{eq2}),
are invariant under the substitutions $D \rightarrow -D$ and $h \rightarrow -h$,
so we will consider only $D>0$ and $h>0$ without loss of generality.

As mentioned before  we consider the AF and F cases of Ising model with DM interaction separately.
However, we only study the non-symmetric pairwise entanglement ($C_{12}$) because the
symmetric one ($C_{13}$) shows the same results qualitatively.
\begin{itemize}
\item AF case ($J>0$)\\
For zero temperature ($T=0$), the ground state is an entangled one
and its concurrence is given by
\bea \no
   C_{12}^{AF}(T=0,h\neq0)=\frac{2D}{q},
\eea
which is plotted in Fig.(\ref{fig13}) and denoted by $T=0$.
The thermal concurrence is nonzero for $T<T_{c}(D,h)$ which
is shown for different temperatures and $h=2.0$ versus $D$ in Fig.(\ref{fig13}).
We have observed that the the critical temperature decreases with increasing of the
magnetic field. Moreover, the DM interaction creates entanglement
in the system. For zero temperature, the onset of DM interaction
leads to nonzero concurrence while for finite temperature $D$
should be greater than a critical value to have nonzero
concurrence. The concurrence is plotted versus $h$ for different
values of temperature and $D=2.0$ in Fig.(\ref{fig15}). This figure shows that for low
temperature the concurrence increases with the increase of
magnetic field initially to reach a maximum and  then decreases
gradually for larger magnetic field, whereas for mid-range
temperature the magnetic field reduces the concurrence slowly.
In other words, for low temperatures field induces entanglement in a
3-qubit system of Ising model with DM interaction.

\begin{figure}
\begin{center}
\includegraphics[width=7cm]{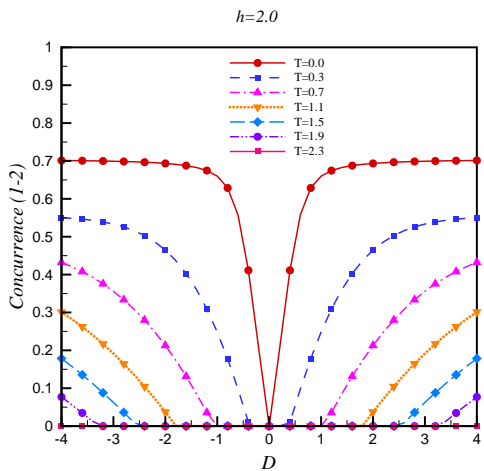}
\caption{(color online) The variations of non-symmetric pairwise
concurrence versus $D$ for different values of $T$ in AF case
($J=1$), $h=2.0$.} \label{fig13}
\end{center}
\end{figure}

\begin{figure}
\begin{center}
\includegraphics[width=7cm]{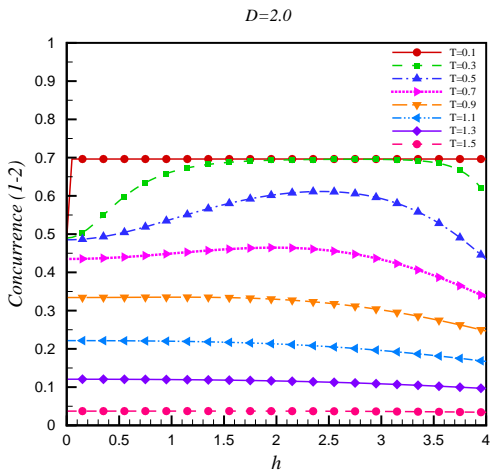}
\caption{(color online) The concurrence of non-symmetric pairwise
as a function of magnetic field ($h$) for different values of $T$
in AF case ($J=1$), $D=2.0$.} \label{fig15}
\end{center}
\end{figure}

\item F case ($J<0$)\\
For $D\leq\sqrt{\big[(3+2h)^{2}-1\big]/8}$, the
entanglement of the ground state is zero while it is nonzero for
$D\geq\sqrt{\big[(3+2h)^{2}-1\big]/8}$. In the entangled region
the concurrence is
\bea \no
   C_{12}^{F}(T=0,h\neq0)=\frac{2D}{q}.
\eea
For non-zero temperature and $D<D_{c}(T, h)$ the concurrence is
zero as shown in Fig.(\ref{fig17}) for $h=2.0$. The increment of DM interaction and the magnetic field
induce entanglement in this system as far as $T<T_{c}(D,h)$.
The critical value of DM coupling ($D_c(T, h)$) is increased with the increment of
magnetic field. Moreover, we have observed that the amount of concurrence is
enhanced for higher magnetic fields.  More
importantly, all figures reveal that the system can be entangled
in a region which is not entangled at $T=0$ by the effect of
thermal fluctuations.

\end{itemize}

\begin{figure}
\begin{center}
\includegraphics[width=7cm]{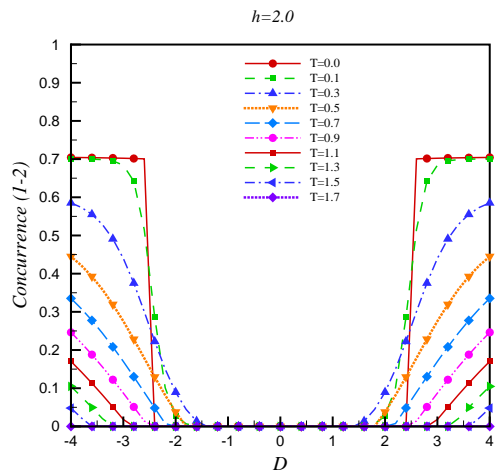}
\caption{(color online) The concurrence of non-symmetric pairwise
against $D$ for different values of $T$ for F case ($J=-1$),
$h=2.0$.} \label{fig17}
\end{center}
\end{figure}


\begin{figure}
\begin{center}
\includegraphics[width=7cm]{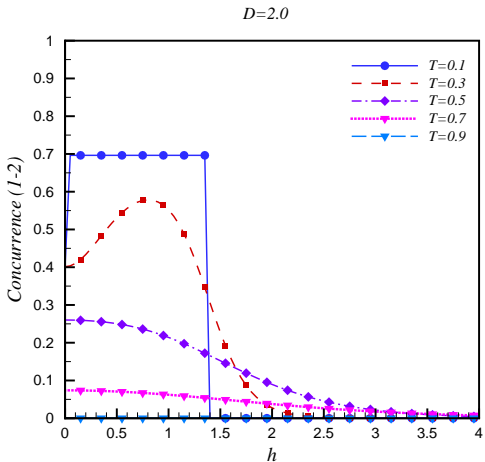}
\caption{(color online) The concurrence of non-symmetric pairwise
is plotted versus $h$ for different values of $T$ in F case
($J=-1$), $D=2.0$.} \label{fig19}
\end{center}
\end{figure}


\section{Summery and Conclusions \label{conclusion}}

The ground state and thermal entanglement of symmetric and
non-symmetric pairwise in three-qubit $XXZ$ and Ising models in the
presence of DM interaction has been investigated. we have studied
the influence of DM and anisotropy parameters on the concurrence
of these models.

The DM interaction and anisotropy are the efficient control
parameters of entanglement. For the positive anisotropy region (AF
case of IDM), the symmetric pairwise is disentangled in finite and zero
temperature. For non-symmetric pairwise,
increasing of anisotropy enhances the ground state and thermal
entanglement of the system to its maximum value and then decreases
it slowly. In this region the DM interaction has a different
effect on the concurrence. For $0<\Delta<1$, the increase of DM
coupling decreases the ground state entanglement and concurrence at low
temperature but at high temperature DM interaction
enhances thermal concurrence. For $\Delta\geq1$, increment of DM
coupling enhances the ground state entanglement and low
temperature concurrence to a maximum value and then decreases it
gradually but it has only an increment effect on the concurrence
at hight temperature.

In the negative anisotropy region (F case of IDM) there is a
threshold for anisotropy and $D$ parameters above which both symmetric
and non-symmetric pairwise are entangled. Decreasing the anisotropy
increases the ground state entanglement of symmetric pairwise but
decreases the ground state entanglement of non-symmetric pairwise.
However, lowering the anisotropy increases the thermal entanglement
of symmetric pairwise up to a maximum value and then decreases
gradually to be vanished, but it decreases the non-symmetric pairwise
thermal concurrence to be eliminated. In symmetric pairwise the increment
of DM coupling decreases the ground state entanglement but enhances
the thermal entanglement up to the maximum value and then
decreases it. For non-symmetric pairwise the influence of DM
interaction on the ground state and thermal concurrence is
incremental.

The most noteworthy  result occurs at $T=0$ where the 3-qubit
ground state entanglement of the systems shows the fingerprint of
quantum phase transition for an infinite size system. For negative
anisotropy case of both symmetric and non symmetric pairwise,
level crossing of the ground state and first excited state occurs
at $\Delta=-\sqrt{(1+D^{2}})$ ($D=1$ for F
case of IDM) under which the concurrence is zero and jumps to a
nonzero value for $\Delta>-\sqrt{(1+D^{2}})$ ($D<1$). For positive
anisotropy case of non-symmetric pairwise, the symmetry breaking
of the ground state occurs similar to the negative anisotropy case
without level crossing at $\Delta=\sqrt{(1+D^{2}})$ (D=1 for AF
case of IDM) and the concurrence has a maximum value due to the
maximum quantum fluctuations. At these points the global
$U(1)\times Z_{2}$ symmetry of the Hamiltonian is changed to local
hidden $SU(2)$ symmetry\cite{Jafari3,Jafari4}.

We have probed the influence of magnetic field in 3-qubit
IDM model. In non-symmetric pairwise and at low
temperature the magnetic field induces the entanglement to the
system which get a maximum for finite field. Obviously for high
magnetic field the model enters a paramagnetic phase which is
disentangled. For middle range of temperature the entanglement is
reduced by adding the magnetic field, whereas for high
temperature, there are many states populated and its reduction is
very tiny. It is related to the thermal mixing of states which is
a source of entanglement in the system. For the ferromagnetic
case, the variations of entanglement is faster than the AF one.

For symmetric pairwise in the AF case the magnetic field can
induce entanglement in a non-entangled system ($h=0$) and the
entanglement enhances with increasing of the DM interaction. The
entanglement is increased with magnetic field to reach a maximum
value while further increment of magnetic field decreases the
entanglement which gradually disappears. This happens very quickly
for F case and the critical temperature where the entanglement is
vanishing is lower than the AF case. The optimal mixing of all
eigenstates in the system leads the maximum value.

More generally, the entanglement properties of a finite system depends
on the number of qubits as discussed in the text.

\begin{acknowledgments}
The authors would like to acknowledge M. Kargarian and
M. F. Miri for useful discussions and comments.
This work was supported in part by the Center of Excellence in
Complex Systems and Condensed Matter (www.cscm.ir).

\end{acknowledgments}


\section*{References}
\begin{thebibliography}{99}

\bibitem{Schrodinger}
E. Schr\"{o}dinger, Proc. Camb. Pill. Soc. \textbf{31}, 555
(1935).

\bibitem{Einstein}
A. Einstein, B. Podolsky, and N. Rosen, Phys. Rev. \textbf{47},
777 (1935).

\bibitem{Bell}
J. S. Bell, Physics \textbf{1}, 195 (1964).

\bibitem{Bennet1}
C. H. Bennet and D. P. Divincenzo, Nature \textbf{404}, 247
(2000).

\bibitem{Gu}
Shi-Jian Gu, Shu-Sa Deng, You-Quan Li, and Hai-Qing Lin, Phys. Rev. Lett \textbf{93}, 086402 (2004);
S. J. Gu, G. S. Tian, and H. Q. Lin, New J. Phys. \textbf{8}, 61 (2006);


\bibitem{Larsson}
Daniel Larsson and Henrik Johannesson, Phys. Rev. Lett. \textbf{95}, 196406 (2005).

\bibitem{Legeza}
\"{O}. Legeza and J. S\'{o}lyom, Phys. Rev. Lett. \textbf{96}, 116401 (2006).

\bibitem{Nielsen}
M. A. Nielsen, Ph.D thesis, Universuty of New Mexico, 1998,
quant-ph/0011036.

\bibitem{Wang1}
X. Wang, Phys. Rev. A \textbf{64}, 012313 (2001).

\bibitem{Kamta}
G. L. Kamta and A. F. Starace, Phys. Rev. Lett \textbf{88}, 107901
(2002).

\bibitem{OConnor}
K. M. Oconnor and W. K. Wootters, Phys. Rev. A \textbf{63}, 052302
(2001).

\bibitem{Sun}
Y. Sun, Y. Chen, and H. Chen, Phys. Rev. A \textbf{68}, 044301
(2003).

\bibitem{Khveshchenko}
D. V. Khveshchenko, Phys. Rev. B \textbf{68}, 193307 (2003).


\bibitem{Zhang1}
G. F. Zhang and S. S. Li, Phys. Rev. A \textbf{72}, 034302 (2005).

\bibitem{Jafari}
R. Jafari, M. Kargarian, A. Langari, M. Siahatgar, Phys. Rev. B
\textbf{78}, 214414 (2008).

\bibitem{kargarian}
M. Kargarian, R. Jafari, A. Langari, Phys. Rev. A \textbf{76},
060304(R) (2007); M. Kargarian, R. Jafari, A. Langari, Phys. Rev.
A \textbf{77}, 032346 (2008).


\bibitem{Nishiyama}
M. Nishiyama, Y. Inada and Guo-qing Zheng, Phys. Rev. Lett
\textbf{98}, 047002 (2007).

\bibitem{Trauzettel}
B. Trauzettel, Denis V. Bulaev, Daniel Loss and Guido Burkard,
Nature Phys. \textbf{3}, 192 (2007).

\bibitem{Porras}
D. Porras,and J. I. Cirac, Phys. Rev. Lett \textbf{92}, 207901
(2004).

\bibitem{Loss}
Daniel Loss, and David P. DiVincenzo1, Phys. Rev. A \textbf{57}, 120 (1998);
Guido Burkard, and Daniel Loss, and David P. DiVincenzo, Phys. Rev. B \textbf{59}, 2070 (1999).

\bibitem{Kane}
B. E. Kane, Nature (London) 393, 133 (1998).

\bibitem{Imamog-lu}
A. Imamog lu, D. D. Awschalom, G. Burkard, D. P. DiVincenzo, D. Loss, M. Sherwin, and A. Small, Phys. Rev. Lett.
\textbf{83}, 4204 (1999).

\bibitem{Zheng}
Shi-Biao Zheng, and Guang-Can Guo, Phys. Rev. Lett. \textbf{85}, 2392 (2000).

\bibitem{Anders}
Anders S{\o}rensen, and Klaus M{\o}lmer, Phys. Rev. Lett.
\textbf{83}, 2274 (1999).


\bibitem{Lidar}
D. A. Lidar, D. Bacon, and K. B. Whaley, Phys. Rev. Lett.
\textbf{82}, 4556 (1999).

\bibitem{Abliz}
 A. Abliz, H. J. Gao, X. C. Xie, Y. S. Wu, and W. M. Liu, Phys. Rev. A \textbf{74},
052105 (2006).

\bibitem{Dominic}
Dominic W. Berry and Mark R. Dowling, Phys. Rev. A \textbf{74},
062301 (2006).

\bibitem{Souza}
A. M. Souza, M. S. Reis, D. O. Soares-Pinto, I. S. Oliveira, and
R. S. Sarthour, Phys. Rev. B 77, 104402 (2008).

\bibitem{Kheirandish}
F. Kheirandish, S. J. Akhtarshenas and H. Mohammadi, Phys. Rev. A
\textbf{77}, 042309 (2008).

\bibitem{Li}
D. C. Li and Z. L. Cao, Eur. Phys. J. D. {\bf 50}, 207 (2008).

\bibitem{Zhang2}
G. F. Zhang, Phys. Rev. A \textbf{75}, 034304 (2007); G. F. Zhang,
J. Phys.: Conddense. Matter \textbf{19}, 456205 (2007)

\bibitem{Chuang}
Da-Chuang Li, Xian-Ping Wang, Zhuo-Liang Cao, J. Phys.: Conddense.
Matter \textbf{20}, 325229 (2008).

\bibitem{Gurkan}
Z. N. Gurkan and O. K. Pashaev, e-print arXiv:quant-ph/0705.0679
and arXiv:quant-ph/0804.0710.

\bibitem{Wang2}
X. G. Wang, Phys. Lett. A, \textbf{281}, 101 (2001).

\bibitem{Dzyaloshinskii}
I. Dzyaloshinskii, J. Phys. Chem. Solids \textbf{4}, 241 (1958).

\bibitem{Moriya}
T. Moriya, Phys. Rev \textbf{120}, 91 (1960).

\bibitem{Dur}
W. D\"{u}r, G. Vidal, and J. I. Cirac, Phys. Rev. A \textbf{62},
062314 (2000).

\bibitem{Coffman}
V. Coffman, J. Kundu, W. K. Wootters, Phys. Rev. A \textbf{61},
052306 (2000).

\bibitem{Brun}
T. A. Brun and O. Cohen, Phys. Lett. A \textbf{281}, 88 (2001).

\bibitem{Acin}
A. Ac\'{i}n, A. Andrianov, L. Costa, E. Jan\'{e}, J. I. Latorre,
and R. Tarrach, Phys. Rev. Lett \textbf{85}, 1560 (2000).

\bibitem{Rajagopal}
A. K. Rajagopal and R. W. Rendell, Phys. Rev. A \textbf{65},
032328 (2002).


\bibitem{Karlsson}
A. Karlsson and M. Bourennane, Phys. Rev. A \textbf{58}, 4394
(1998).

\bibitem{Hao}
J. C. Hao, C. F. Li, and G.C. Guo, Phys. Rev. A \textbf{63},
054301 (2001).

\bibitem{Brub}
D. Bru{\ss}, D. P. DiVincenzo, A. Ekert, C. A. Fuchs, C.
Macchiavello, and J. A. Smolin, Phys. Rev. A \textbf{57}, 2368
(1998).

\bibitem{Dender1}
D. C. Dender, P. R. Hammar,D. H. Reich, C. Broholm , and G. Aeppli
Phys. Rev. Lett \textbf{79}, 1750 (1997).


\bibitem{Kohgi}
M. Kohgi, K. Iwasa, J. Mignot, B. Fak, P. Gegenwart, M. Lang, A.
Ochiai, H. Aoki, and T. Suzuki, Phys. Rev. Lett \textbf{86}, 2439
(2000).

\bibitem{Tsukada}
I. Tsukada, J. T. Takeya, T. Masuda and K. Uchinokura, Phys. Rev.
Lett \textbf{87}, 127203 (2001)

\bibitem{Grande}
b. Grande and Hk. M$\ddot{u}$ller-Buschbaum, Z. Anorg. Allg. Chem
\textbf{417}, 68 (1975).

\bibitem{Greven}
M. Greven, R. J. Birgeneau, Y. Endoh, M. A. Kastner, M. Matsuda, and
G. Shirane, Z. Phys. B \textbf{96}, 465 (1995).



\bibitem{Alcaraz}
F. C. Alcaraz and W. F. Wreszinski, J. Stat. Phys. \textbf{58}, 45
(1990).

\bibitem{Wootters}
W. K. Wootters, Phys. Rev. Lett \textbf{80}, 2245 (1998).

\bibitem{Jafari3}
R. Jafari, A. Langari, arXiv:0804.4579.

\bibitem{Jafari4}
M. Kargarian, R. Jafari, A. Langari, arXiv:0812.18.62.


\end{thebibliography}
\end{document}